\documentclass[proceedings]{JHEP}
\usepackage{epsfig}                   % please use epsfig for figures
\def\LY{Lee-Yang}

\def\hhc{{{\hat h}_{\rm crit}}}
\newcommand{\be}{\begin{equation}}
\newcommand{\ee}{\end{equation}}

\renewcommand{\bar}{\overline}
\newcommand{\xtra}[1]{{.}}
\renewcommand{\xtra}[1]{{, \tt hep-th/#1.}}
\newcommand{\xxtra}[1]{{, \tt hep-th/#1}}
\newcommand{\xtrac}[1]{{.}}
\renewcommand{\xtrac}[1]{{, \tt cond-mat/#1.}}

\newcommand{\mathematica}[1]{{}}
\newcommand{\bp}[3]{{\ensuremath{ \phi_{#1}^{(#2#3)}}}}

\newcommand{\newL}{R}

\newcommand{\ds}{\displaystyle}
\newcommand{\mm}[2]{{\vphantom{\vbox to 6mm{}}}}
\newcommand{\fract}[2]{{\textstyle\frac{#1}{#2}}}

\newcommand{\ri}{\right}
\newcommand{\ep}{\varepsilon}
\newcommand{\lf}{\left}
\newcommand{\te}{\theta}

\newcommand{\beqcol}{\begin{array}{rcl}}
\newcommand{\eeqcol}{\end{array}}

\newcommand{\CK}{{\cal K}}
\newcommand\eq{\begin{equation}}
\newcommand\en{\end{equation}}
\newcommand\bea{\begin{eqnarray}}
\newcommand\eea{\end{eqnarray}}
\newcommand\nn{\nonumber}

\newcommand{\One}{{\hbox{{\rm 1{\hbox to 1.5pt{\hss\rm1}}}}}}
\renewcommand{\One}{{\mathbb 1}}
\renewcommand{\One}{{\rm 1\!\!1}}

\newcommand{\JP}[1]{J.\ Phys.\ {\bf #1}}

\newcommand{\iintd}{\int^{\infty}_{-\infty}\! d\theta}

\newcommand{\ket}[1]{|#1\rangle}

\newcommand{\prtial}{\frac{\partial}{\partial\theta}}

\newcommand{\opnup}[1]{\renewcommand{\\}{\\[50 pt]}}
\renewcommand{\bar}{\overline}

\newcommand\Eblk{{\cal E}_{\rm bulk}}

\newcommand{\D}{{{\rm d}}}
\newcommand{\vev}[1]{\langle\,#1\,\rangle}

\newcommand{\cH}{{\cal H}}

\newcommand{\IJMP}[1]{Int.\ J.\ Mod.\ Phys.\ {\bf #1}}

\newcommand{\NP}[1]{Nucl.\ Phys.\ {\bf #1}}
\newcommand{\PL}[1]{Phys.\ Lett.\ {\bf #1}}

\newcommand{\AlBZ}{Al.B.~Zamolodchikov}
\conference{Nonperturbative Quantum Effects 2000 }

\title{ Finite size effects in  perturbed boundary conformal field theories}

\author{
P.Dorey, M.Pillin, A.Pocklington, I.Runkel,
R.Tateo, G.M.T.Watts  \\

SPhT Saclay, 91191 Gif-sur-Yvette, France, and
Dept. Math. Sciences, University of Durham, Durham DH1 3LE, England (PED) \\

ETAS, PTS-A, Borsigstr.10, D-70469 Stuttgart, Germany (MP)\\

IFT/UNESP, Instituto de Fisica Teorica, 01405-900, Sao Paulo - SP, Brasil (AP)\\

King's College London, Strand, London WC2R 2LS, England (IR and GMTW)\\

UVA, Inst. voor Theoretische Fysica, 1018 XE Amsterdam, The Netherlands (RT)\\

E-mail:\email{ p.e.dorey@durham.ac.uk, 
mathias.pillin@etas.de,
andrew@ift.unesp.br, 
ingo@lpthe.jussieu.fr, 
tateo@wins.uva.nl, 
gmtw@mth.kcl.ac.uk
}
}
\abstract{          
We discuss the finite-size properties of a simple integrable quantum 
field theory in 1+1 dimensions with non-trivial boundary conditions. 
Novel off-critical identities between cylinder partition functions of 
models with differing boundary conditions are derived. 
(Talk given by RT at the TMR conference `Nonperturbative Quantum Effects 2000')
\centerline{
 SPhT-T00/135, DTP/00189, P.083/2000, KCL-MTH-00-53, ITFA 00-14;}
\centerline{ PRHEP-tmr2000/035, {\tt hepth/0010278}} \\[-10pt]
}
\keywords{Finite size effects, Integrable models, Thermodynamic Bethe ansatz}
\begin{document}
\section{Introduction}
A fair amount of  work has been devoted in recent years to the  
investigation of 
non-perturbative phenomena in integrable quantum field theories. 
Besides the fact that such systems might be used as a
testing ground for  general ideas in quantum field theory, such as 
RG flows and dualities,  
the wide success of the topic  is probably due to its relevance
in condensed matter physics. 
More recently, variants of such systems defined on non-trivial geometries
have been considered~\cite{GZ}.
This is a topic which has an even larger set of
potential applications.  
In  condensed matter physics, for example, these include the study of 
Kondo-type systems and the fractional quantum 
Hall effect (see 
\cite{Sa1,Sa2} for detailed reviews of recent
results in these research areas). 
String theorists~\cite{HKM} have also shown 
a certain interest in the subject, and  
surprisingly  connections  with other 
relevant pieces of modern mathematics 
and physics~\cite{BLZ1,DTb} have been 
discovered.

In this note we shall sketch  results more  
extensively presented in a series of 
collaborative works~\cite{Us1,Us2,Us3,Us4}  
concerning  
a simple  interacting quantum field theory 
confined on a strip-type geometry. 

The  ${\cal M}_{2,5}$ model is 
perhaps  the  simplest nontrivial
rational Conformal 
Field Theory 
(CFT). It has central charge $c=-22/5$, and can be identified with 
the non-trivial   RG  fixed point of the
($T<T_c$) Ising model in a strong  purely imaginary magnetic field. 
This fixed point coincides with the  accumulation point of the Lee-Yang 
zeros and  the corresponding conformally invariant theory 
is known as the \LY\ CFT.  

The 
model  contains only two irreducible 
representations of the Virasoro algebra.
These have weights $0$ and $-1/5$,
and the corresponding bulk primary fields are the identity $\One$, and a 
scalar field $\varphi$ of scaling dimension $x_\varphi\,{=}\,{-}2/5$.
There are  two
conformally-invariant boundary conditions  denoted 
by $\One$ and $\Phi$ and 
three relevant boundary fields
interpolating  pairs of boundary conditions.  
They all have conformal weight $-1/5$. 
Two of these fields  (denoted $\psi$ and $\psi^\dagger$) interpolate different
boundary conditions, while the third ($\phi$) lives on 
the $\Phi$ boundary:
\eq
  \psi \equiv \bp {-1/5}\One\Phi
\;,\;\;\;
  \psi^\dagger \equiv \bp {-1/5}\Phi\One   
\;,\;\;\;
  \phi
\equiv
  \bp {-1/5}\Phi\Phi
\;.
\en
\section{The perturbed  CFT}
We shall discuss a 
perturbation of this CFT, the scaling \LY~model.
On a cylinder of width $R$ and circumference $L$, this
has the action
\bea
\ds{{\cal A}_{BLY}
= {\cal A}_{BCFT}+\lambda\int_0^R \D x\int_0^L \D y\,\varphi(x,y)
}\nn \\
\ds{+h_l\int_0^L \D y\,\phi(0,y)
+h_r\int_0^L \D y\,\phi(R,y)\,,~~~~~} 
\eea 
with coordinates $0\,{\leq}\,x\,{\leq}\,\newL$ 
across the cylinder, and 
$0\,{\leq}\,y\,{<} L$
around it. The parameter $\lambda$ determines the bulk mass 
and we have allowed the possibility of boundary-perturbing fields on the 
left and right 
ends of the cylinder, with couplings $h_l$ and $h_r$.   
${\cal A}_{BCFT}$ is the conformally-invariant action  on the cylinder
with (conformal) boundary conditions $(\Phi,\Phi)$.
We shall also  consider
$(\One,\Phi)$ and $(\One,\One)$ boundary conditions which have a 
similar expression for ${\cal A}_{BLY}$, but lack 
one or 
both perturbing boundary fields.

There are two possible  Hamiltonian 
descriptions of the cylinder  partition function. 
In the  so-called L-channel representation the r{\^o}le of time 
is taken by $L$:
\bea
\ds{
  Z_{\alpha\beta}
=
{\rm Tr}_{\cH_{(\alpha,\beta)}} e^{-LH_{\alpha\beta}^{\rm strip}(M,R)}
{}~~~~~~~~~~~~~~}\label{rrchan}  \\[3pt]
\ds{=
 \sum_{E_n\in\,{\rm spec}(H_{\alpha\beta}^{\rm strip})}\, \,
 e^{ - L E_n^{\rm strip}(M,R)}
}\;,~~~ 
\nn
\eea
while  in  the R-channel representation  the r{\^o}le of time is taken 
by $R$:
\bea
&&\ds{
  Z_{\alpha\beta} 
=   \vev{\alpha|
  \,e^{-RH^{\rm circ}(M,L)}\,
  |\beta}~~~~~~~~~~~} \label{llchan}   \\[3pt]
&&\qquad\ds{
=\sum_{E_n\in\,{\rm spec}(H^{\rm circ})}
 \!\!
\fract{\vev{\alpha|\psi_n}\vev{\psi_n|\beta}}{\vev{\psi_n|\psi_n} }
\, e^{ - R E_n^{\rm circ}(M,L) }.}\nn
\eea
In equation 
(\ref{llchan}) we have used boundary 
states 
$\ket{\alpha}$,$\ket{\beta}$ $\in \{\ket{\One},\ket{\Phi} \}$ 
and the  eigenbasis $\{ \ket{\psi_n} \}$
of the Hamiltonian $H^{\rm circ}\,,$ which propagates states living on a 
circle  of   circumference $L$.
By contrast, $H_{\alpha\beta}^{\rm strip}$ in (\ref{rrchan})
propagates states along a
strip of width $R$, and
acts on the Hilbert space $\cH_{(\alpha,\beta)}$
of states on an interval
with boundary conditions $\alpha$ and $\beta$ imposed on the
two ends.
\vskip 8pt
\[
\begin{array}{c}
\refstepcounter{figure}
\label{fig:lchan}
\epsfxsize=.70\linewidth
\epsfbox{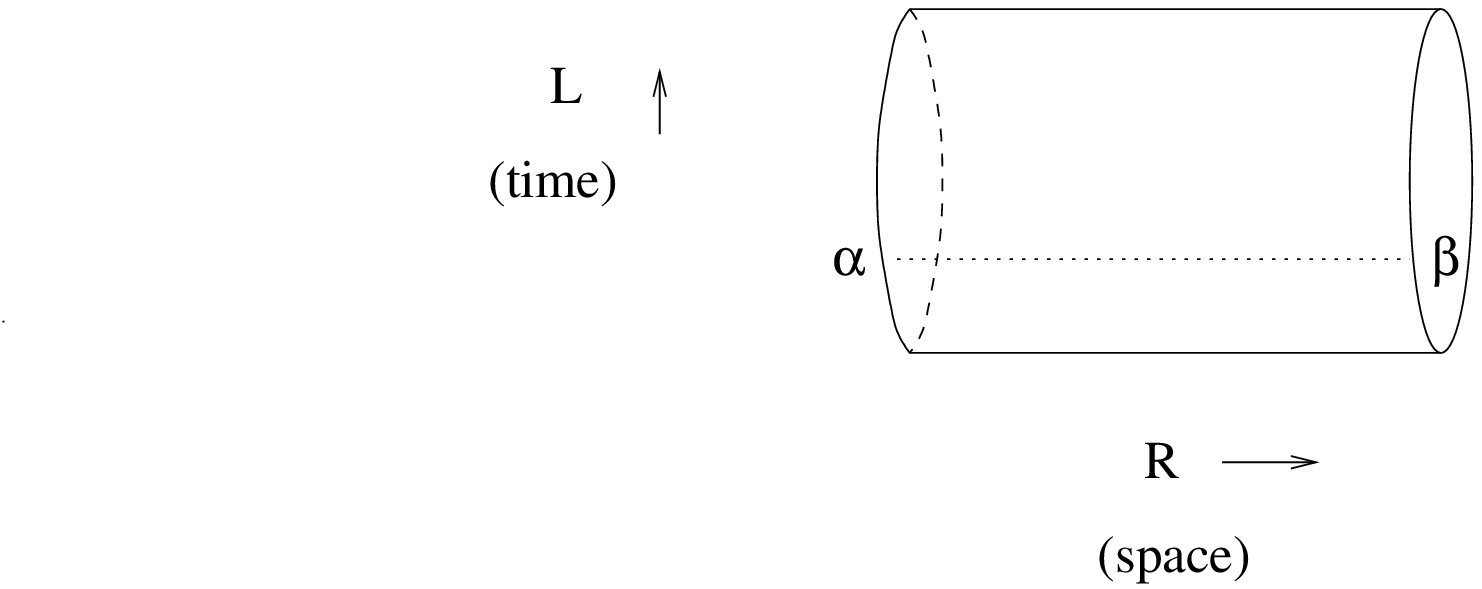}
\\
\parbox{.99\linewidth}{\footnotesize\raggedright
Figure \ref{fig:lchan}:~The L-channel decomposition: states $\ket{\chi_n}$ live on the dotted
line segment across the cylinder.
}
\end{array}
\] 
\[
\begin{array}{c}
\refstepcounter{figure}
\label{fig:rchan}
\epsfxsize=.70\linewidth
\epsfbox{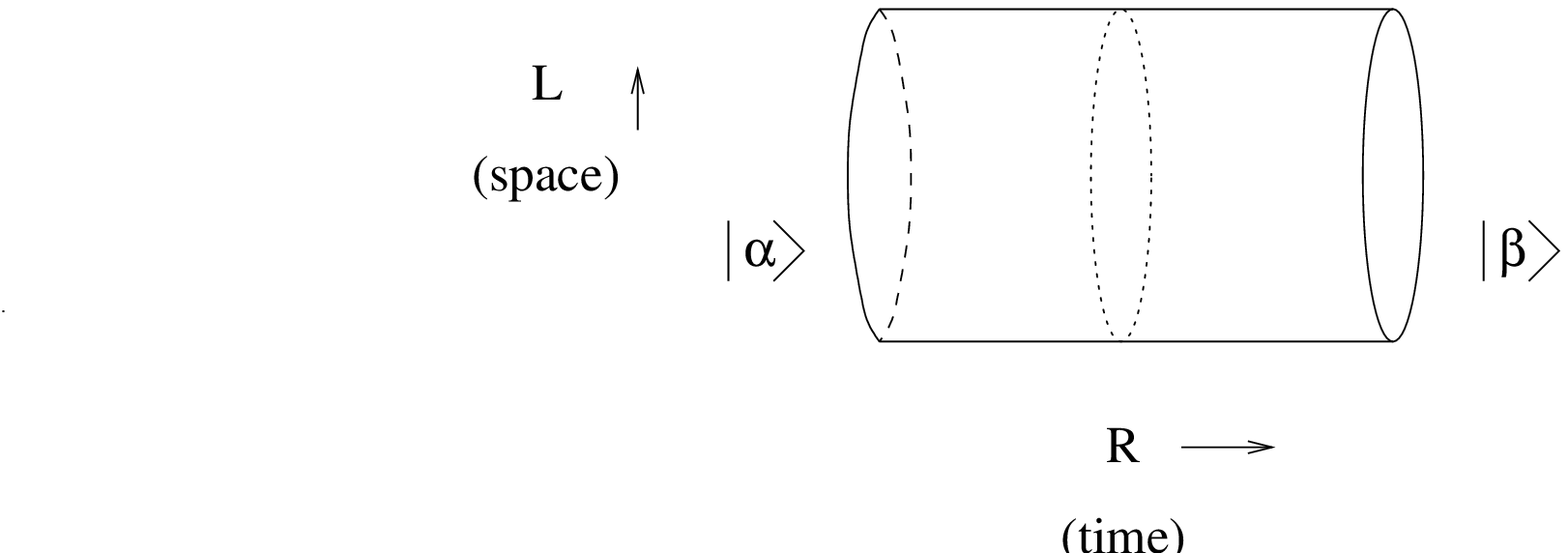}
\\
\parbox{.99\linewidth}{\footnotesize\raggedright
Figure \ref{fig:rchan}:~The R-channel decomposition: states $\ket{\psi_n}$ live on the dotted
circle around the cylinder.
}
\end{array}
\] 

\noindent
The two decompositions are illustrated in figures \ref{fig:lchan} and
\ref{fig:rchan}.

Conformal field theory provides the following useful 
representations for ${H}^{\rm circ}$ and ${H}^{\rm strip}$:
\eq
{H}^{\rm circ}
= 
\frac{2 \pi}{L} 
    \Big(
  L_0 + \bar{L}_0- \fract{c}{12}+
    \lambda  
  \left| \fract{L}{2 \pi} \right|^\fract{12}{5}
    \int\limits_{\theta = 0}^{2\pi} \varphi(e^{i\theta}) \,\D\theta  
  \Big)
\;, 
\label{h2}
\en
\bea
\ds{{H}^{\rm strip}
=
\frac{\pi}{\newL} 
    \Big(
  L_0-\fract{c}{24}
+
    \lambda 
    \left| \fract{\newL}{\pi} \right|^\fract{12}{5}
    \!\!\int\limits_{\theta = 0}^\pi \!\!
    \varphi(e^{i\theta})
    \,\D\theta} ~~\nn \\
\ds{+
     h_l
    \left| \fract{\newL}{\pi} \right|^\fract{6}{5}
    \phi_l(-1)
+
    h_r
    \left| \fract{\newL}{\pi} \right|^\fract{6}{5}
    \phi_r(1)
  \Big)~
}~.
\label{h1}
\eea 
The truncated conformal space approximation, or TCSA, 
gives numerical 
estimates for the low-lying eigenvalues of the 
Hamiltonians~(\ref{h2})~\cite{YZ} and (\ref{h1})~\cite{Us1} 
via a  diagonalisation of the matrix  
$\vev{i|H|j}$, where $\ket{i}$ and $\ket{j}$ are  
states in a finite-dimensional 
subspace of the relevant CFT Hilbert space. 

If the bulk coupling $\lambda$ is set to zero in (\ref{h1}),
then a purely boundary flow can be studied 
using the TCSA~\cite{Us1}.  
\[
\begin{array}{c}
\refstepcounter{figure}
\label{Bflow}
\epsfxsize=.70\linewidth
\epsfbox{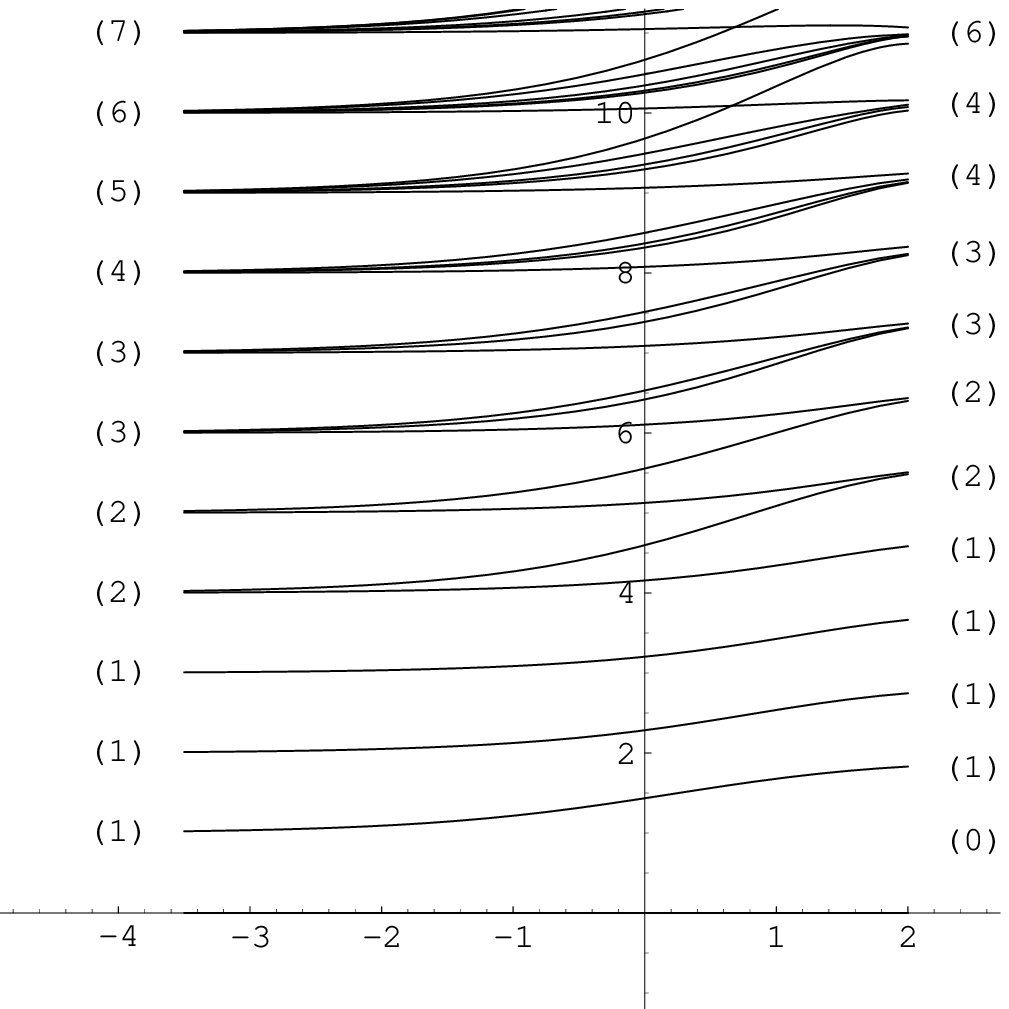}
\\
\parbox{.99\linewidth}{\footnotesize\raggedright
Figure \ref{Bflow}:~The boundary spectral  flow.
}
\end{array}
\] 
The 
gaps for the excited state scaling functions 
$F_n(h^{5/6}R)=R E_n^{\rm strip}(h,R)/\pi$
are plotted in 
figure~\ref{Bflow} 
as a function of $\log(h^{6/5}R)$ for the
model on a strip with $(\One,\Phi(h))$ boundary conditions. 
Note how the multiplicities (written in parentheses) reorganise 
themselves
to give a smooth flow between the conformal $(\One,\Phi)$ and 
$(\One,\One)$ spectra, as
encoded in the $\chi_{-1/5}$ and $\chi_0$
Virasoro characters of ${\cal M}_{2,5}$\,.

In more complicated models the structure of the boundary flows can be much
richer; see, for example, 
Gerard Watts' talk at this conference~\cite{GRW}.
\section{The scaling \LY\ model}
Turning now to non-zero values of $\lambda$,
the scaling \LY\  model can also be  described by a 
massive scattering theory consisting
of a single self-conjugate particle species ($A=\bar{A}$) 
with $2 \rightarrow 2$ S--matrix \cite{CM}
\[
S_{AA}^{AA}(\te)= S(\te)
=  -(1)(2)
\;,\;
  (x)
=  \fract{\sinh \lf( \fract{\te}{2} +\fract{i \pi x}{6} \ri)
     }{
    \sinh \lf( \fract{\te}{2} -\fract{i \pi x}{6} \ri)}
\;.
\]
The physical strip pole at $\theta= 2 \pi i/3$ in $S(\theta)$ 
corresponds to the on-shell tree diagram involving a non-vanishing 
``$\varphi^3$'' coupling
represented in figure~\ref{smapole}, 
while 
the pole at $\theta=i \pi/3$ describes the same process  seen from the 
crossed channel.
The mass $M$ of the particle is
related to the bulk perturbation parameter $\lambda$ by 
$M  = \kappa \lambda^{5/12}$. The  exact value of the constant 
$\kappa$ was found in~\cite{Alkappa}.
\[
\begin{array}{cc}
\refstepcounter{figure}
\label{sma}
\epsfxsize=.50\linewidth
\epsfbox{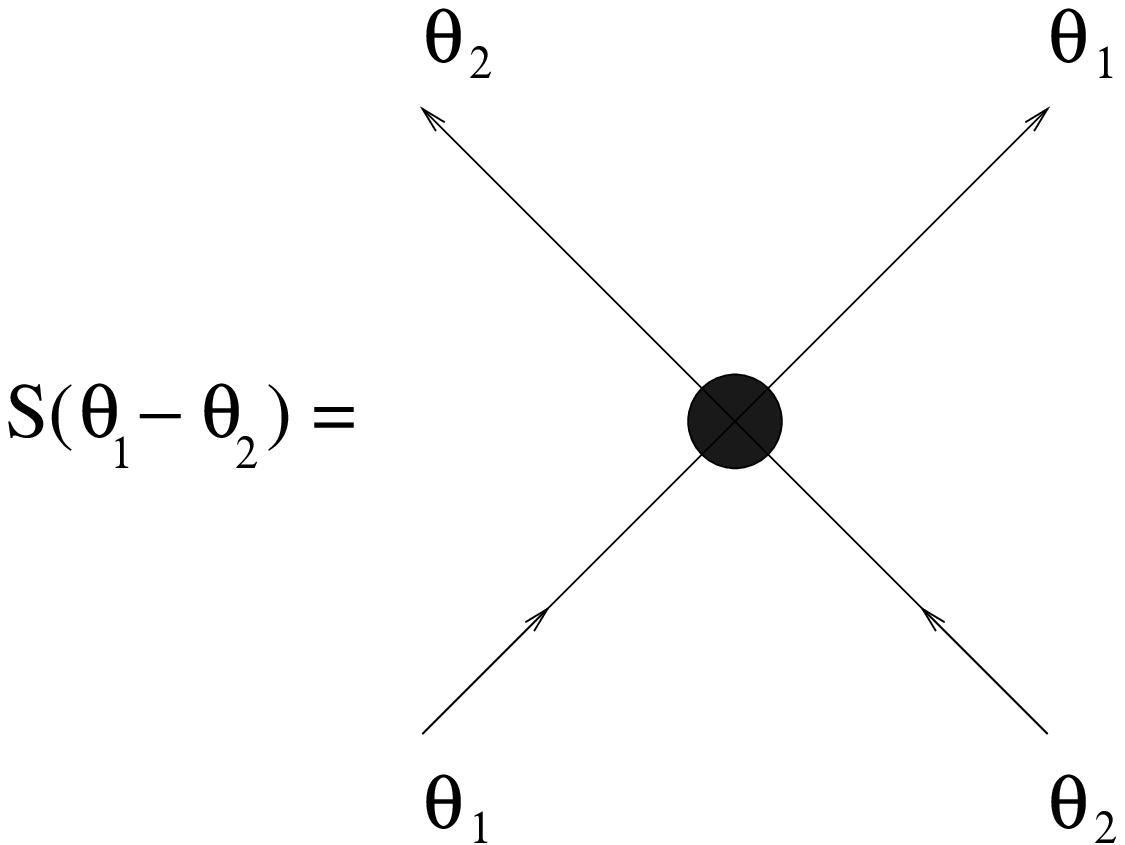}{~~~~} % 3d
&
\refstepcounter{figure}
\label{smapole}
\epsfxsize=.50\linewidth
{}\epsfbox{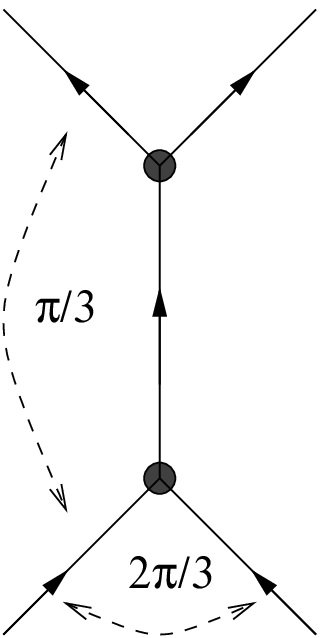}{~~~}
\\
\parbox[t]{.47\linewidth}{\footnotesize\raggedright%
Figure \ref{sma}:~The the $2 \rightarrow 2$  S-matrix.
}~
&
\parbox[t]{.47\linewidth}{\footnotesize\raggedright%
Figure \ref{smapole}:~The ``$\varphi^3$'' property.
}
\end{array}
\]
Placing an impenetrable wall at the coordinate position $x=0$,
the theory must be supplemented by  a reflection factor    
describing how the particle bounces off the boundary.    
The reflection amplitudes corresponding to the two integrable boundary
conditions $\One$ and $\Phi$ were found in \cite{Us1}. They  are
\eq
R_\One(\te)= 
\lf(\fract{1}{2}\ri)\lf(\fract{3}{2}\ri)
  \lf(\fract{4}{2}\ri)^{-1}~,
\en
and $R_{\Phi(h)}(\te)=R_b(\te)$, where
\eq
  R_b(\te)
= R_{\One}(\te)
  \lf( S(\te+i \pi \fract{b+3}{6})
  S(\te-i \pi \fract{b+3}{6}) \ri)^{-1}\,
\en
and 
\eq
h(b) = -|\hhc| M^{6/5} \sin( \pi (b+1/2)/5)~.
\en
The exact value
of $\hhc$ was determined in~\cite{Us3}.
Notice that the reflection factors for the $\One$ and $\Phi(h(0))$
boundaries
are identical. However the 
particle -- boundary interaction properties differ:
the
type $\One$ boundary is particle repelling while $\Phi(h(0))$ 
is attractive~\cite{Us2,Us4}. 
We shall consider cases where $b$ is real and  restrict
$b$ to the range $[-3,3]$. The physical poles of 
$R_\One(\te)$ and  $R_b(\te)$  at 
$\theta=i \pi/6$ and 
$\theta=i \pi/2$ can be explained by postulating 
a non-vanishing boundary-particle coupling, while the poles at 
$-i \theta=V_{10}= \pi(b+1)/6$ and at $-i\theta=V_{20}=\pi (b-1)/6$ in
$R_b(\te)$  lie in the physical strip for $b \in [-1,2]$
and  $b \in [1,2]$ respectively. 
It is natural to associate the latter poles to 
boundary bound states.
The energies $e_1$ and $e_2$ of these states are given by
$e_j -e_0 =M \cos(V_{j0})$, with $e_0$ being the energy of the boundary 
ground-state. Notice  that the difference  $e_1-e_0$ is negative 
for   $b \in [2,3]$ and the first boundary bound state becomes, in this range,
the true vacuum state.
The other poles in the reflection 
amplitude $R_b$  
also have
field theoretic explanations, but only when 
a boundary analogue
of the
Coleman-Thun mechanism is invoked~\cite{Us2}.
Further discussion of this aspect of boundary scattering can be
found in \cite{GG,MD}. 
\[
\begin{array}{cc}
\refstepcounter{figure}
\label{rmat}
\epsfxsize=.45\linewidth
\epsfbox{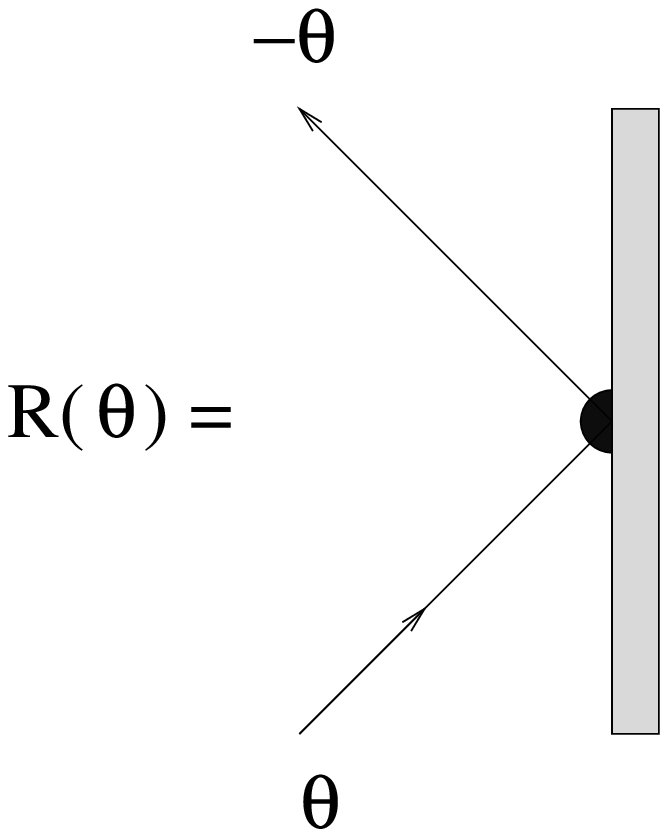}{~~~} % 3d
  &
\refstepcounter{figure}
\label{rmatpole}
\epsfxsize=.45\linewidth
{}\epsfbox{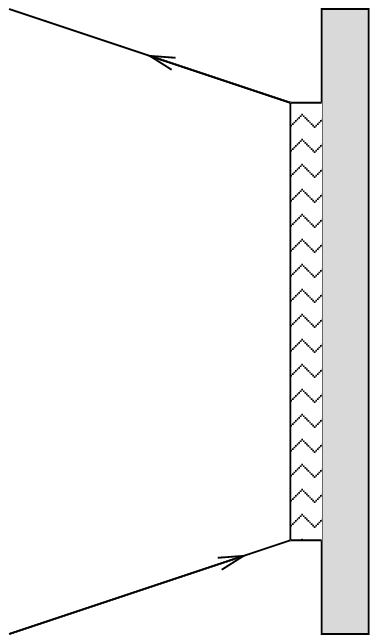}{~~~}
\\
\parbox[t]{.47\linewidth}{\footnotesize\raggedright%
Figure \ref{rmat}:~The reflection matrix.
}~
  &
\!\!\parbox[t]{.47\linewidth}{\footnotesize\raggedright%
Figure \ref{rmatpole}:~Formation of a boundary bound state.
}
\end{array}
\]
\section{The thermodynamic Bethe ansatz}
The Thermodynamic 
Bethe Ansatz (TBA)~\cite{Zb} has proved to be  a powerful 
tool 
in the study of the ground-state energy  of  
integrable quantum field theories on a infinite cylinder. 
More recently, a  variant of the method (the BTBA)
has been proposed to describe boundary 
situations~\cite{LMSS}. In~\cite{Us1}, a detailed analysis
of these equations 
for the case of the \LY~model 
was performed. The analysis 
revealed, for the $(\One,\Phi)$ boundary
conditions, that the equations of~\cite{LMSS} correctly describe 
the  ground-state
energy in the range $b \in [-3,-1]$ but need to be 
modified for  $b \in [-1,2]$ by the inclusion of  extra `active 
singularity'
terms (see \cite{BLZe,DT1,FMQR,PF,Us1} 
and eq.~(\ref{kermit}) below).
Moreover, excited state energies 
$E^{\rm strip}_{n}$ can also
be computed, using similar generalisations. 
For the scaling
\LY\ model the analysis of~\cite{LMSS,Us1} led to a
non-linear integral equation for a single function $\ep(\theta)$:
\eq
  \ep(\theta)
= \nu(\theta)
+ \sum_p \log {S(\theta - \theta_p) \over {S(\theta - \bar{\theta}_p)}}
 - \CK {*}L(\theta) \,,
\label{kermit}
\en
plus an associated set of equations for the finite (and
possibly empty)
set $\{\theta_p,\bar{\theta}_p\}$ of 
`active'
singularities:
$\exp(\ep(\theta_p))= \exp( \ep(\bar{\theta}_p))=-1$ $\forall p$.
In (\ref{kermit}), $L(\theta)=\log\bigl(1{+}e^{-\ep(\theta)}\bigr)$,
the symbol ``$*$'' indicates  the standard convolution,
and
\eq
\nu(\theta)=2\,RM\cosh\theta-\log\lambda_{\alpha \beta }(\theta) \,,
\en
\eq
\lambda_{\alpha\beta}(\theta)=R_{\alpha}(i \fract{\pi}{2}-\theta)
R_{\beta}(i \fract{\pi}{2}+\theta)~,
\en 
\eq
\CK(\theta)=-i\prtial\log S(\theta).
\en
The number of active singularities depends on the particular 
energy level considered and, as mentioned above, 
for some pairs of boundary conditions
on the strip it is nonzero even for the ground state~\cite{Us1}.
The solution to~(\ref{kermit}) for
a given value of $r=RM$ 
determines a  function $c_n(r)$:
\bea
\ds{
c_n(r)\,=\,\frac{6}{\pi^2}\iintd\, r\cosh\theta L(\theta) 
~~~~~~~~~~~}
\nn \\
\ds{+ i {12 r \over \pi}
\sum_p (\sinh \theta_p -\sinh\bar{\theta}_p) \, ,} 
\label{piggy}
\eea
in terms of which
$
   E_n^{\rm strip}(M,R)
= \Eblk R
+  f_{\alpha}+f_{\beta}
- \frac{\pi}{24 R}c_n(r)
$, 
where the $f$'s  are  
 R-independent contributions to  the energy from
the  boundaries and $\Eblk$ is the bulk energy per unit length. Exact 
expressions for these quantities can be found, for example, in~\cite{Us4}. 
In figure~\ref{comp} we compare the numerical 
diagonalisation (TCSA) of $H^{\rm strip}$ with boundary 
condition $(\One,\Phi|_{b=-1.5})$
with the numerical solution of the BTBA.
No boundary bound states are present.
\vskip -30pt
\[
\begin{array}{c}
\refstepcounter{figure}
\label{comp}
\epsfxsize=.90\linewidth
\epsfbox{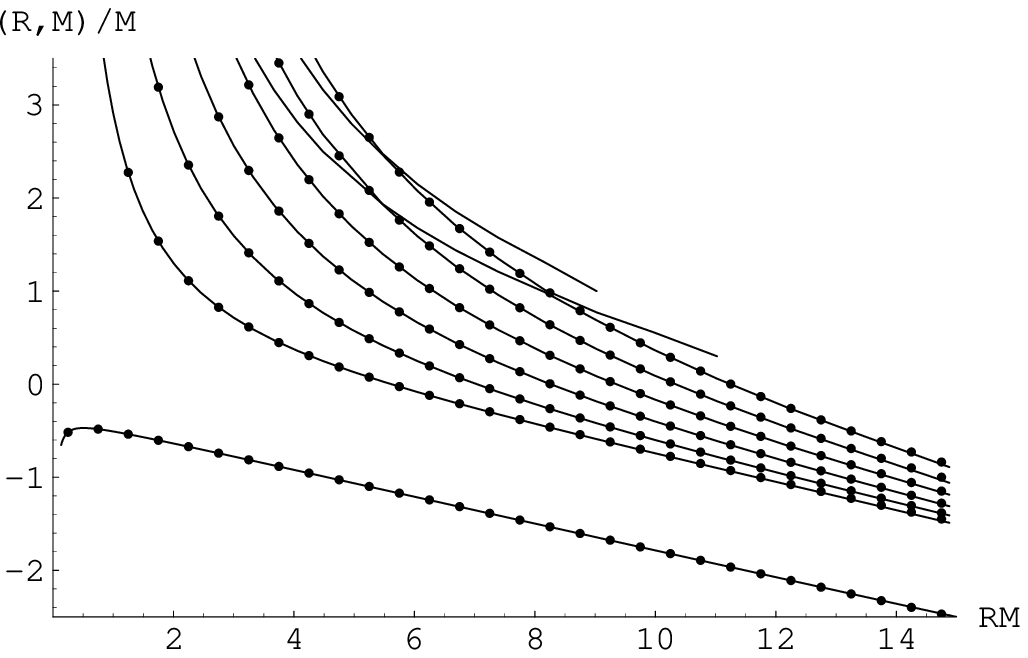}
\\[-18pt]
\parbox{.80\linewidth}{\footnotesize\raggedright
Figure \ref{comp}:~TCSA (lines) versus  BTBA (dots).
}
\end{array}
\] 
Figure~\ref{bbs} shows instead the situation for $(\One,\Phi|_{b=0.8})$:  
the first excited state now corresponds to a boundary bound state 
with an energy gap tending, as $R \rightarrow \infty$, to 
$\Delta e=e_1-e_0$ $=\cos(V_{10}|_{b=0.8})$.
\vskip -30pt
\[
\begin{array}{c}
\refstepcounter{figure}
\label{bbs}
\epsfxsize=.90\linewidth
\epsfbox{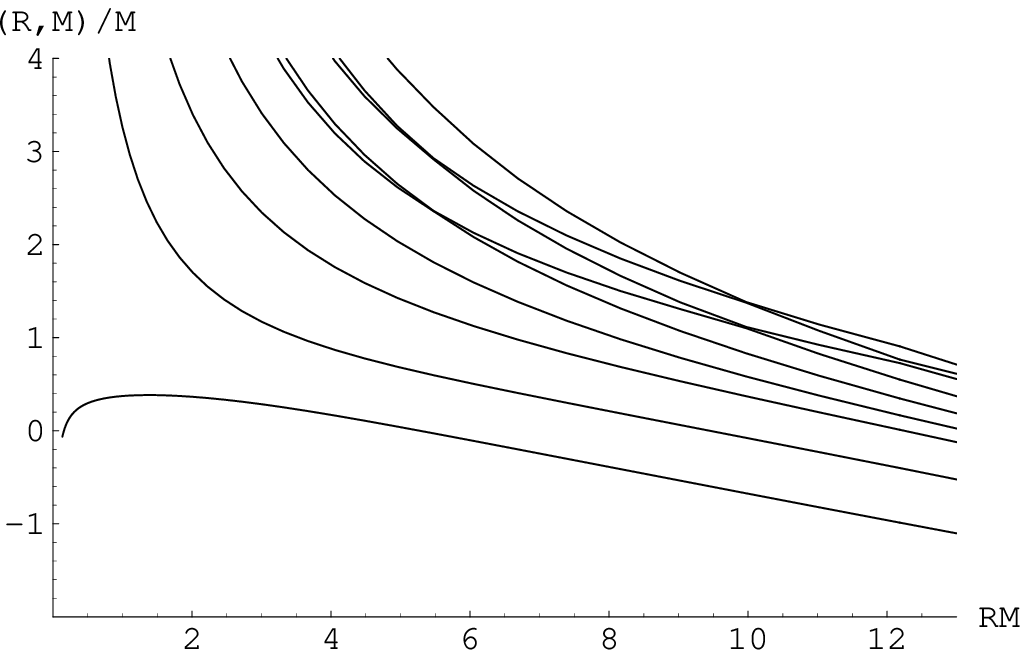}
\\[-18pt]
\parbox{.99\linewidth}{\footnotesize\raggedright
Figure \ref{bbs}:~TCSA result: the first boundary bound state.
}
\end{array}
\] 
\vskip -10pt
Note that the finite-size data has allowed us to confirm the existence
of the boundary bound state, inferred at the end of the last section
by a completely independent calculation based on the pole structure
of the proposed reflection factor.

\section{The partition function identities}
Let us now consider the R-channel decomposition of the partition 
function~(\ref{llchan}). 
The following identification was made in \cite{BLZ1,Us3} for the 
coefficients of the  weights $\exp(-RE^{\rm circ}_{n})$:
\eq
\vev{\Phi(h(b))|  \psi_n}=
Y_n(i \pi \fract{b+3}{6}) \vev{\One|\psi_n} \,,
\label{g=y}
\en
where $ Y_n(\theta) = e^{\ep_n(\theta)}$,
and
$\ep_n(\theta)$ is the solution of the $n^{\rm th}$ excited-state 
TBA equation  with periodic boundary conditions. These equations are
simply recovered by   setting $\nu(\theta)= LM \cosh\theta$ and $r=LM$ in 
(\ref{kermit}) and (\ref{piggy}).
Relation  (\ref{g=y}) can be then  used to write the 
partition function as
\bea
\ds{
Z_{\Phi(h(b_r)),\Phi(h(b_l))} = 
\sum_n \Big( 
\fract{\vev{\One|\psi_n}\vev{\psi_n|\One} 
}{
\vev{\psi_n|\psi_n}} ~~~~~~} \nn \\
\ds{
Y_n(i \pi \fract{b_r+3}{6})
Y_n(i \pi \fract{b_l+3}{6})
e^{-R \,{E^{\rm circ}_{n}(M,L)}} \Big)~.~\nn
}
\eea
Recalling
that the  Y's satisfy the  functional relation~\cite{AlZY}
\footnote{Note, the
`ground-state' $Y_0$  at $\lambda{=}0$
coincides with the Stokes multiplier of a Schr{\"o}dinger equation
with $x^3$ 
potential~\cite{DTb}} 
\eq
Y_n(\theta+ i {\pi/3} )Y_n(\theta- i {\pi/3} )  
=1 + Y_n(\theta)~,
\en
we quickly obtain  the following  identity 
\eq
Z_{\Phi(h(-b)),\Phi(h(b))}=Z_{\One,\One}+
Z_{\One,\Phi(h(2-b))}\, .
\label{zid}
\en
Eq.~(\ref{zid}), being valid for arbitrary 
$R$ and $L$,  
is  equivalent to the following 
relation between the spectra of models on strips of equal widths
but different boundary conditions:
\bea
\ds{
\{  E^{\rm strip}_n \}_{\Phi(h(-b)),\Phi(h(b))} =~~~~~~~~~~~~~~~~} \nn \\ 
\ds{\{  E^{\rm strip}_n \}_{\One,\One}
\cup
\{  E^{\rm strip}_n   \}_{\One,\Phi(h(2-b))}} \, .
\label{eid} 
\eea
This rather surprising relationship  has been 
numerically checked  using the
TCSA approach and it can be considered as a first off-critical extension
of identities between conformal partition functions 
provided in, for example, \cite{BPZb}. 
We shall now discuss a simple application of~(\ref{eid})~\cite{Us4}.

There are regions in which the model with
$~(\Phi(h_l),\Phi(h_r))~$ boundary conditions develops a boundary-induced
vacuum instability: an example situation is represented
in figure~\ref{vin}.
(The dashes correspond to complex-conjugate pairs of spectral lines.)

\[
\begin{array}{c}
\refstepcounter{figure}
\label{vin}
\epsfxsize=.99\linewidth\
\epsfysize=.65\linewidth\
\epsfbox{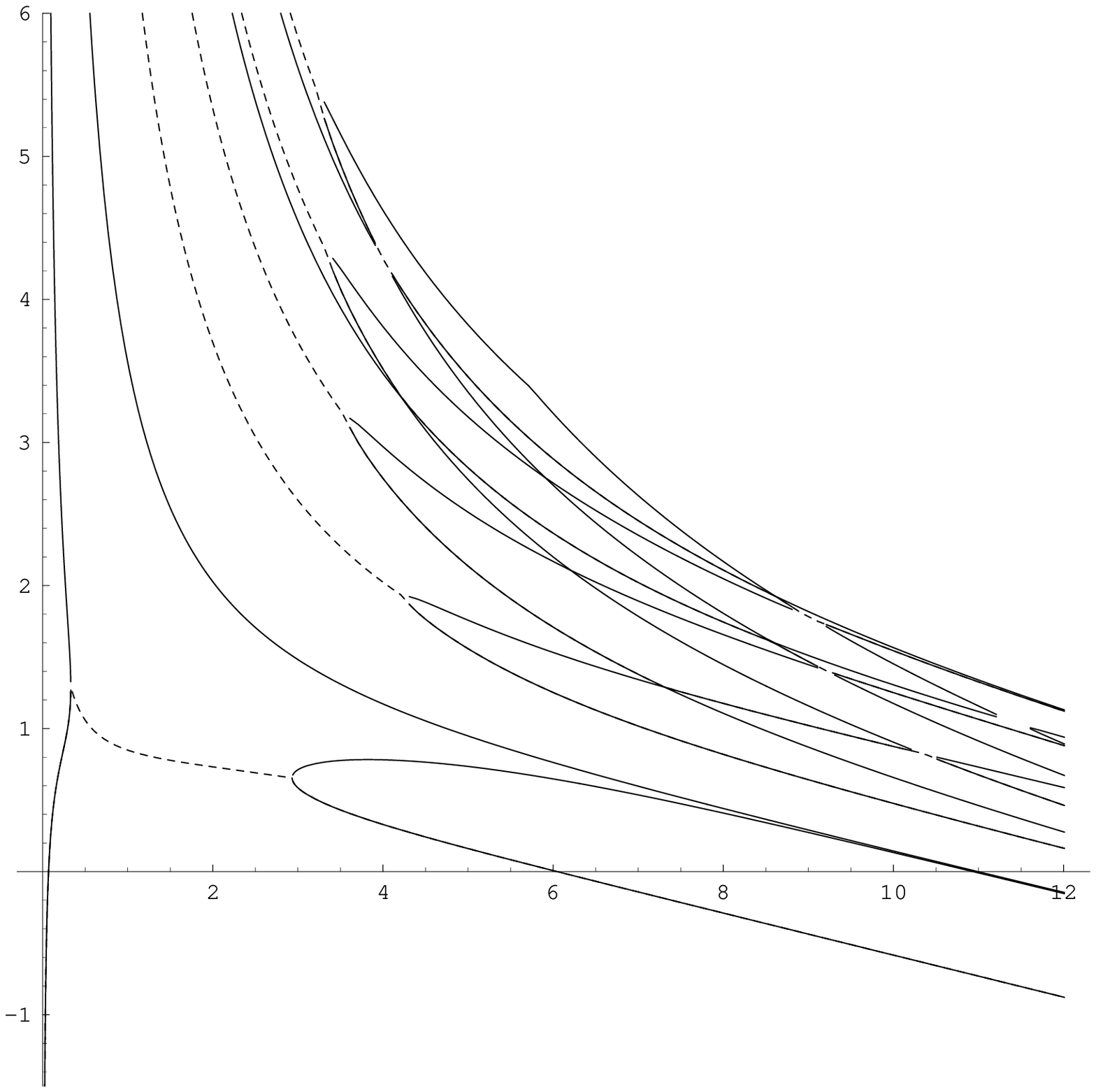}
\\
\parbox{.99\linewidth}{\footnotesize\raggedright
Figure \ref{vin}:~Example of vacuum instability ($b_l=b_r=\fract{1}{2}$).
(Axes as on figures~{\protect \ref{comp}} and~{\protect \ref{bbs}}.)
}
\end{array}
\] 

These regions are shaded on the $(h_l,h_r)$ plane in figure \ref{ellipse}. The
inner dashed line delimits the values of $h_l$ and $h_r$ covered for
real values of $b_l$ and $b_r$, while on the ellipse the identity
(\ref{eid})  holds.
On the ellipse, in the  small $R$ region,  the  energy levels  
$E_0|_{\One,\Phi(h(2-b))}$ 
and  $E_0|_{\One,\One}$   
correspond,  respectively, to the ground state  and  to the 
first excited state
in the  $(\Phi(h(-b))$,$\Phi(h(b)))$ model. In the opposite, large-$R$, limit, 
the energy gap 
$\Delta E=E_0|_{\One,\One}- E_0|_{\One,\Phi(h(2-b))}$ tends for $b>0$ to  
$f_{\One}-f_{\Phi(h(2-b)} = \sin {(b-2) \pi/6}$ and becomes
negative for $|b|<2$. This simple fact signals the presence 
of a level crossing at some intermediate value of $R$. The RHS of (\ref{eid})
prohibits the mixing of the two states and ensures that the crossing will be 
exact. The set of $b$-values with $\Delta E<0$ corresponds to the portion of the ellipse on 
figure~\ref{ellipse} which touches the shaded region.
Once the line $\Delta b= b_l+b_r=0$ is left, the identity (\ref{eid}) can 
no longer be invoked and the level crossing it lost. As $\Delta b$ decreases
the two levels repel, while in the opposite direction the boundary fields
are stronger suggesting  the presence of a vacuum instability. 
Finally in the region  
$h_l < -|\hat{h}_{\rm crit}| M^{6/5}$ ($h_r <-|\hat{h}_{\rm crit}| M^{6/5}$) the reflection factor $R_{b(h_l)}(\theta)$ is not 
a pure phase for real $\theta$ and, consequently,  
the vacuum is already unstable in the 
infinite volume~\cite{Us1}. 
Shading the region(s) 
within which, 
for at least one value of the strip width, the model exhibits a 
 boundary-induced instability, we end up with the phase-diagram
represented in  figure~\ref{ellipse}. The conjectured scenario has been 
checked using  the TCSA method. 
\[
\begin{array}{c}
\refstepcounter{figure}
\label{ellipse}
\epsfxsize=.60\linewidth\
\epsfbox{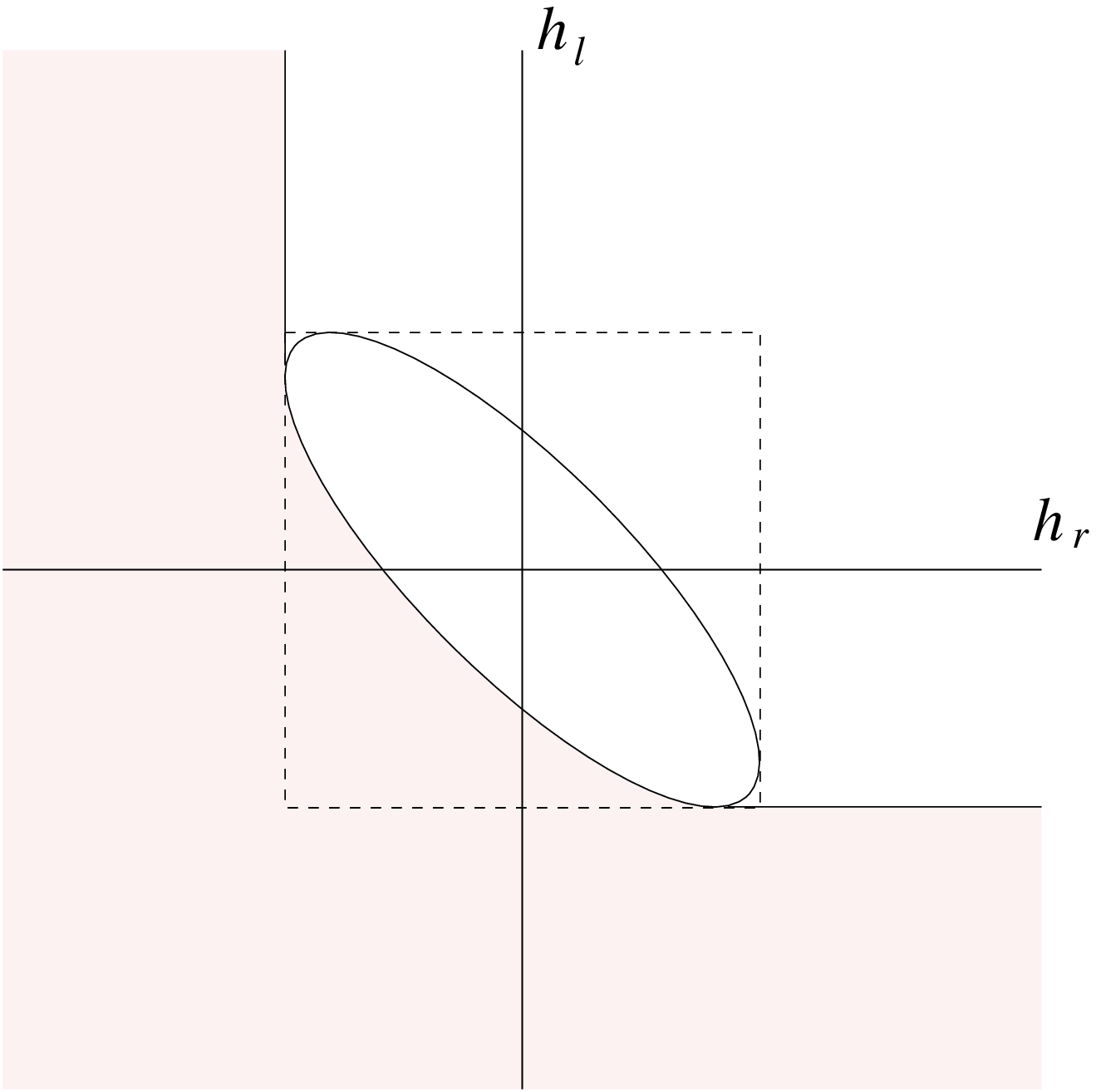}
\\
\parbox{.99\linewidth}{\footnotesize\raggedright
Figure \ref{ellipse}:~The phase diagram. 
}
\end{array}
\] 
%
%%%%
%
\section{Conclusions}

In  this note we have given a brief account of  results
on boundary quantum field theory obtained in the last few
years 
in a series of collaborative papers.
Some of the observed phenomena, such as 
 the partition function
identities, are expected to be common features of
integrable boundary models defined 
on a cylinder geometry.

Constraints of space and time prevent us from describing two
other important topics that can be studied using the example of
the boundary scaling Lee-Yang
model. In~\cite{Us3}, a detailed examination
of the flow of the ground-state degeneracy function $g$ was made.
It was found that the TBA equations proposed in~\cite{LMSS} 
do not quite capture the variation of $g$ with changes of the
bulk mass scale, though if this scale is held fixed, they
correctly describe its behaviour as a function of the boundary
parameters. 

Then the main theme of the recent work~\cite{Us4}
was  the computation of correlation functions in a semi-infinite
geometry. 
We refer the interested reader to the original papers for a
detailed discussion of these and other issues.

\acknowledgments
RT thanks Denis Bernard and Bernard Julia for the invitation to 
speak at
this conference.
The work was supported in part by a TMR grant of the European
Commission, reference ERBFMRXCT960012.

{\bf Note added:} since this talk was given, an interesting paper \cite{RaAh}  has
appeared, in which various aspects of a supersymmetric generalisation of
the scaling Lee-Yang model in the presence of boundaries are discussed.

%%%%%%%%%%%%%%%%%%%%%%%%%%%%%%%%%%%
%


\begin{thebibliography}{99}
\raggedright
%
\bibitem{GZ} 
S.\,Ghoshal and  A.B.\,Zamolodchikov,
{\em `Boundary S matrix and boundary state in two-dimensional integrable
quantum field theory'},
\IJMP{A9} (1994) 3841-3886, 
erratum \IJMP{A9} (1994) 4353\xtra{9306002}
%
%
\bibitem{Sa1}
H.\,Saleur,
{\em `Lectures on non perturbative 
Field Theory and Quantum Impurity Problems:
part~I'},
Les Houches Summer School ``Topology and Geometry in Physics'', 
July 1998\xtrac{9812110} 
%
\bibitem{Sa2}
H.\,Saleur,
{\em `Lectures 
on non perturbative Field Theory and Quantum Impurity Problems:
part~II'},
Nato Advanced Study Institute/EC Summer School on ``New Theoretical
Approaches to Strongly Correlated Systems'', Newton Institute, April 
2000\xtrac{0007309}
%
%
\bibitem{HKM}
J.A.\,Harvey, D.\,Kutasov and E.J.\,Martinec,
{\em `On the relevance of tachyons'}\xtra{0003101}
%
%
\bibitem{BLZ1}
V.V.\,Bazhanov, S.L.\,Lukyanov and A.B.\,Zamolodchikov,
{\em `Integrable Structure of Conformal Field Theory, Quantum KdV Theory and 
Thermodynamic Bethe Ansatz',}
Commun. Math. Phys. 177 (1996) 381\xtra{9412229}
%
%
\bibitem{DTb}
P.\,Dorey and R.\,Tateo,
{\em `Anharmonic oscillators, the thermodynamic Bethe ansatz and  nonlinear 
 integral equations',} J.Phys. A32 (1999) L419\xxtra{9812211};\\
P.\,Dorey and R.\,Tateo,
{\em `On the relation between Stokes multipliers and the T-Q system of 
  conformal field theory',}
\NP{B563} (1999) 573\xxtra{9906219}; \\  
P.\,Dorey, C.\,Dunning and R.\,Tateo,
 {\em `Ordinary Differential Equations and
Integrable Models',}
(talk by PED at this conference), 
 PRHEP-tmr2000/034\xtra{0010148}
%
%
\bibitem{Us1}
P.\,Dorey, A.\,Pocklington, R.\,Tateo and G.M.T.\,Watts,
{\em
`TBA and TCSA with boundaries and excited states',}
\NP{B525} (1998) 641\xtra{9712197}
%
%
\bibitem{Us2}
P.\,Dorey, R.\,Tateo and G.M.T.\,Watts,
{\em
`Generalisations of the Coleman-Thun mechanism and boundary reflection
factors',}
\PL{B448} (1999) 249\xtra{9810098}
%
%
\bibitem{Us3}
P.\,Dorey, I.\,Runkel, R.\,Tateo and G.M.T.\,Watts,
{\em 
`$g$--function flow in perturbed boundary conformal field theories',} 
\NP{B578} (2000) 85\xtra{9909216}
%
%
\bibitem{Us4}
P.\,Dorey, M.\,Pillin, R.\,Tateo and G.M.T.\,Watts,
{\em `One-point functions in perturbed boundary conformal field 
theories',} \NP{B} (to appear)\xtra{0008039} 
%
%
\bibitem{YZ}
V.P.\,Yurov and  \AlBZ,
{\em `Truncated conformal space approach to the scaling Lee-Yang model',}
\IJMP{A5} (1990) 3221.
%
\bibitem{GRW}
K.\,Graham, I.\,Runkel and  G.M.T.\,Watts,
{\em `Renormalisation group flows of boundary theories',}
(talk by GMTW at this conference),
PRHEP-tmr2000/040\xtra{0010082} 
%
\bibitem{CM}
J.L.\,Cardy and G.\,Mussardo,
{\em `S matrix of the Yang-Lee edge singularity in two dimensions',}  
\PL{B225} (1989) 275.
%
%
\bibitem{Alkappa}
\AlBZ,
{\em `Mass scale in sine-Gordon model and its reductions',}
Int. J. Mod. Phys {\bf A10} (1995) 1125.
%
\bibitem{GG}
G.W.\,Delius and G.M.\,Gandenberger,
{\em `Particle Reflection Amplitudes in $a_n^{(1)}$ Toda Field Theories',}
\NP{B554} (1999) 325\xtra{9904002}
%
\bibitem{MD}
P.\,Mattsson and  P.\,Dorey,
{\em `Boundary spectrum in the sine-Gordon model with Dirichlet boundary 
conditions'}
\xtra{0008071}
%
\bibitem{Zb}
\AlBZ, 
{\it
`Thermodynamic Bethe Ansatz in Relativistic Models. Scaling
 3-state Potts and Lee-Yang Models',} 
\NP{B342} (1990) 695.
%
%
\bibitem{LMSS}
A.\,Leclair, G.\,Mussardo, H.\,Saleur and  S.\,Skorik,
{\em `Boundary energy and boundary states in integrable quantum field
theories',}
\NP{B453} (1995) 581\xtra{9503227}
%
\bibitem{BLZe}
V.V.\,Bazhanov, S.L.\,Lukyanov and  A.B.\,Zamolodchikov,
{\em `Integrable 
quantum field theories in finite volume: excited state energies',}
\NP{B489} (1997) 487\xtra{9607099}
%
\bibitem{DT1}
P.\,Dorey and R.\,Tateo,
{\em `Excited states by analytic continuation of TBA equations',}
\NP{B482} (1996) 639\xxtra{9607167};\\
P.\,Dorey and R.\,Tateo,
{\em `Excited states in some simple perturbed conformal field theories',}
\NP{B489} (1998) 575\xtra{9706140}
%
\bibitem{FMQR}
D.\,Fioravanti, A.\,Mariottini, E.\,Quattrini and  F.\,Ravanini, 
{\em `Excited state Destri-De Vega equation for Sine-Gordon and
restricted Sine-Gordon Models',}
\PL{B390} (1997) 243\xtra{9608091} 
%
\bibitem{PF}
P.\,Fendley,
{\em `Excited-state energies and supersymmetric indices',}
Adv.Theor.Math.Phys.{\bf 1} (1998) 210\xtra{9706161}
%
\bibitem{AlZY}
\AlBZ,
{\em `On the thermodynamic Bethe ansatz equations for the reflectionless
ADE scattering theories',}
\PL{B253} (1991) 391.
%
%
\bibitem{BPZb}
R.E.\,Behrend, P.A.\,Pearce and  J.-B.\,Zuber,
{\em `Integrable boundaries, conformal boundary conditions and A-D-E
fusion rules',}
\JP{A31} (1998) L763\xxtra{9807142}.
%
\bibitem{RaAh}
C.\, Ahn and R.I.\, Nepomechie,
{\em `The scaling supersymmetric Yang-Lee model with boundary'}\xtra{0009250} 
%
%
\end{thebibliography}
\end{document}